\documentclass[prl,preprintnumbers,amsmath,amssymb,tightenlines,twocolumn,superscriptaddress]{revtex4-2}
\usepackage{graphicx}
\usepackage{psfrag}
\usepackage{color}
\usepackage[utf8]{inputenc}
\usepackage{ifthen}
\usepackage{amsmath}
\usepackage{listings}
\usepackage{placeins}
\usepackage{float}
\usepackage{physics}
\usepackage{comment}
\usepackage[dvipsnames]{xcolor}
\usepackage{soul}
\usepackage[inline]{enumitem}
\usepackage{braket}
\usepackage{hyperref}
\usepackage{xcolor}

\usepackage[overload]{empheq}

\usepackage{graphicx}
\usepackage{psfrag}

\usepackage{pgffor}
\usepackage{dcolumn}

\usepackage{array,ragged2e}
\usepackage[utf8]{inputenc}
\usepackage{ifthen}
\usepackage{bbold}
\usepackage{listings}
\usepackage{placeins}

\newcommand{\E}[1][\empty]{
	\ifthenelse{\equal{#1}{\empty}}
	{\mathbb{E}}
	{\mathbb{E}\left( #1 \right)}
}
\renewcommand{\exp}[1][\empty]{
	\ifthenelse{\equal{#1}{\empty}}
	{\mathrm{exp}}
	{\mathrm{e}^{#1}}
}

\newcommand{\psit}[1][\empty]{%
	\ifthenelse{\equal{#1}{\empty}}
	{\psi_t}
	{\psi_t^{(#1)}}
}

\newcommand{\npsit}[1][\empty]{%
	\ifthenelse{\equal{#1}{\empty}}
	{\tilde\psi_t}
	{\tilde\psi_t^{(#1)}}
}

\newcommand{\SI}{Supplemental Material}

\newcommand{\Bn}{\mathbf{n}}

\newcommand{\Be}{\mathbf{e}}

\definecolor{olive}{RGB}{107,142,35}

\usepackage{ulem}  
\newcommand{\oalex}[1]{{\color{blue}{}}}

\definecolor{olive}{RGB}{107,142,35}

\definecolor{orange}{RGB}{255,139,61}

\graphicspath{{{./}}}

\begin{document}

\title{Non-Markovian Stochastic Schr\"odinger Equation:\\ Matrix Product State Approach to the Hierarchy of Pure States
}

\author{Xing Gao}
\email{gxing@mail.sysu.edu.cn}
\affiliation{School of Materials, Sun Yat-sen University, Shenzhen, Guangdong 518107, China }

\author{Jiajun Ren}
\email{renjj@mail.tsinghua.edu.cn}
\affiliation{MOE Key Laboratory of Organic OptoElectronics and Molecular Engineering, Department of Chemistry, Tsinghua University, Beijing 100084, China }

\author{Alexander Eisfeld}
\affiliation{Max-Planck-Institut f\"ur Physik komplexer Systeme, N\"othnitzer Str.\ 38,
	D-01187 Dresden, Germany }

\author{Zhigang Shuai}
\affiliation{MOE Key Laboratory of Organic OptoElectronics and Molecular Engineering, Department of Chemistry, Tsinghua University, Beijing 100084, China }


\begin{abstract}	 
	 We derive a stochastic hierarchy of matrix product states (HOMPS) for non-Markovian dynamics in open quantum system at finite temperature, which is numerically exact and efficient. 
	 HOMPS is obtained from the recently developed stochastic hierarchy of pure states (HOPS) by expressing HOPS in terms of formal creation and annihilation operators. The resulting stochastic first order differential equation is then formulated in terms of matrix product states and matrix product operators.  In this way the exponential complexity of HOPS can be reduced to  scale polynomial with the number of particles. 
	 The validity and efficiency of HOMPS is demonstrated for the spin-boson model and long chains where each site is coupled to a structured, strongly non-Markovian environment.
\end{abstract}

\maketitle

 Many physical and chemical processes require to take the interaction with environmental degrees of freedoms (DOFs) into account~\cite{breuer2002theory,oliver2008charge}. 
Often some of these DOF couple strongly to the system and exhibit a memory, i.e., they are non-Markovian.
	To handle this challenging situation a variety of approaches have been put forward~\cite{cao2014ttm,tanimura1989time,tanimura2006stochastic,yan2016deom_reivew,makri1995quapi_I,makri1995quapi_II,shi2003GQME,rabani2011GQME,mayer2000mctdh_review,wang2003multilayer,yanshao2016stochastic}.
	One promising approach is the hierarchy of stochastic pure states (HOPS)~\cite{eisfeld2014hierarchy,eisfeld2015hierarchical} which is  a stochastic, wavefunction based open quantum system method, to solve the  non-Markovian quantum state diffusion (NMQSD) equation ~\cite{strunz1997pla,strunz1998pra,strunz1999prl,yu1999perturb,yu2010prl,you2014nmqsd,you2015higher}.
	HOPS has been successfully applied, for example to study energy transfer in small photosynthetic systems~\cite{eisfeld2014hierarchy,zhao2016hierarchy}, or to simulate linear \cite{RiSuMoe15_034115_} and non-linear spectroscopy~\cite{eisfeld2016nonlinearspectra}. 
	HOPS consists of a set of coupled first-order stochastic differential equations.
	For large systems with strong coupling to several distinct environments, HOPS still requires a substantial computational effort because the number of coupled equations grows exponentially with the number of effective environmental modes. 
	\begin{figure}
		\includegraphics[width=8.6 cm]{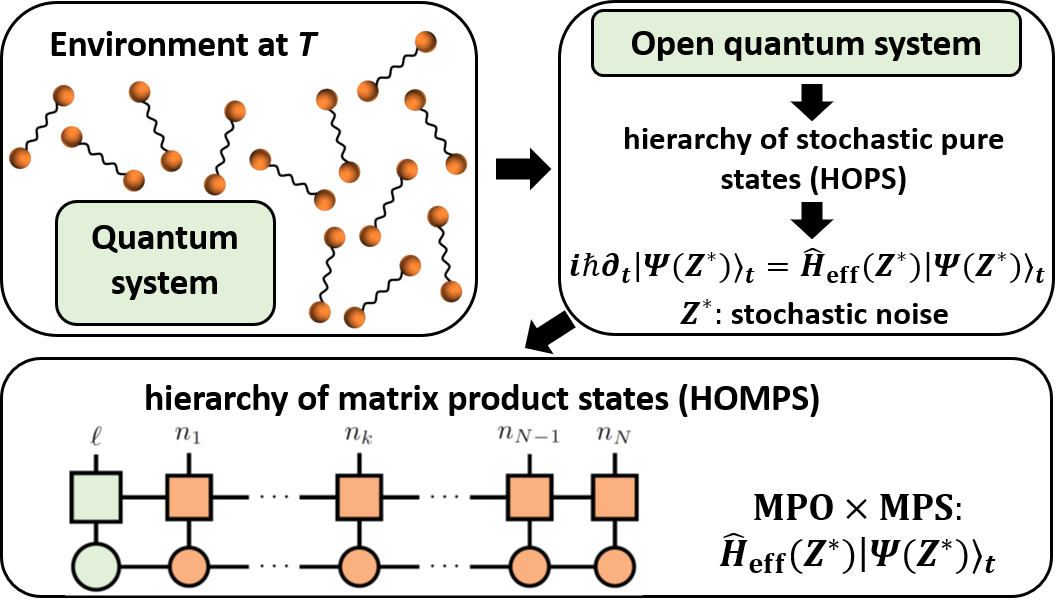}
		\caption{\label{fig1} Our strategy to construct the hierarchy of matrix product states (HOMPS). 
		We consider a quantum system interacting with a thermal environment at temperature $T$ consisting of bosonic modes which are linearly coupled to the quantum system. 
		After tracing out the environmental degrees of freedom, this many-body problem is then treated using the hierarchy of stochastic pure states (HOPS) method. 
		From this, we transform to a stochastic  effective Schr\"odinger type equation  which can be solved efficiently using the MPS/MPO representation.}
	\end{figure}
	
	In this Letter, we show that a substantial reduction of the computational effort can be achieved by formulating HOPS in terms of matrix product states (MPS) and matrix product operators (MPO). 
	The resulting stochastic hierarchy of matrix product states (HOMPS) can be efficiently propagated \cite{schollwock2011density} by modern algorithms that have been used and tested for different problems~\cite{vidaltebd2004,verstraete_tdvp_prl2011,verstraete_tdvp_prb2016,lubich2015time,schollwockl2019time,plenio2010efficient,plenio2019efficient,AlexChin_2018tensor,reiher2020jcprev,ren2018time,shuai2020carrier,mahaibo2019time,Lovett2018efficient,garnet2021openquantum,mctdh_mps2018,shi2018mps,shi2021mpo,raffaele2019heom,pollock2019processtensor,filippov2019processtensor}.
Our procedure is illustrated in Fig.~\ref{fig1} and described in the following.

	\paragraph{The open quantum system:}
		We consider a quantum system coupled linearly to a (infinite) set of harmonic oscillators.
	The total Hamiltonian is written as
	\begin{equation}
		\label{eq:H_tot}
		\hat{H}_{\text{tot}}=\hat{H}_\text{S}+\hat{H}_\text{B}+\hat{H}_{\text{SB}},
	\end{equation}
	with  $\hat{H}_{\text{S}}$,  $\hat{H}_{\text{B}}$, and  $\hat{H}_{\text{SB}}$  describing the system, the bath and the system-bath interaction, respectively. 
	We consider a bath that can consist of several independent parts: $\hat{H}_\text{B}=\sum_{j=1}^J \hat{H}_{\mathrm{B},j}$ with
	$\hat{H}_{\mathrm{B},j}=\sum_\lambda (\frac{\hat{p}_{\lambda,j}^2}{2}+\frac{1}{2}\omega_{\lambda,j}^2 \hat{q}_{\lambda,j}^2)$ where  $\{\hat{p}_{\lambda,j}\}$ and $\{\hat{q}_{\lambda,j}\}$ are the coordinates and momenta of the bath DOFs.  
	The system-bath coupling Hamiltonian is taken as
	\begin{equation}
	\label{eq:H_SB}
		\hat{H}_\text{SB}=\sum_{j=1}^J H_{\text{SB},j}=\sum_j\hat{L}_j \otimes\sum_{\lambda,j} c_{\lambda,j}\hat{q}_{\lambda,j},
	\end{equation} 
	where each system operator $\hat{L}_j$ couples to its own environment.
  	 The interaction strength between system and the $(\lambda,j)$-th mode is quantified by the coefficient $c_{\lambda,j}$. 
  	 It is convenient to define the spectral densities,
  $
  		S_j(\omega)=\frac{\pi}{2}\sum_\lambda\frac{c_{\lambda,j}^2}{\omega_{\lambda,j}}\delta(\omega-\omega_{\lambda,j}),
  $
	which describes the frequency dependent system-bath coupling strength of the $j$-th bath. 
	In the time-domain, the bath correlation function,
	\begin{equation}
		\label{eq:BCF}
		\begin{aligned}
			\alpha_j(t)=&\frac{1}{\pi}\int_0^\infty \!\!\!\!\mathrm{d}\omega S_j(\omega)\big[\coth(\frac{\omega}{2 T})\cos\omega t-i\sin\omega t\big],\\
		\end{aligned}
	\end{equation}
	fully characterizes the influence of the environment at temperature $T$.
    We use the units $\hbar=k_B=1$. 
    
    We are interested in the dynamics of the system which is given by the reduced density matrix
    \begin{equation}
    	\rho(t)=\Tr_\text{B}\{{\rho_\text{tot}(t)}\}.
    \end{equation}
	Here, $\Tr_\text{B}\{\cdots\}$ denotes the trace over all bath DOFs, and $\rho_\text{tot}(t)$ is the total density matrix. 
	In the following, we assume a factorized initial state $\rho_\text{tot}(0)=\rho(0)\otimes \frac{e^{- H_\text{B}/T}}{Z_\text{B}}$ with partition function $Z_\text{B}=\Tr_\text{B}\{e^{- H_\text{B}/T}\}$.
	
	To improve readability, we drop the index $j$ and show explicitly the derivation for a single operator $\hat{L}$, which we take to be Hermitian for simplicity, i.e., $\hat{L}=\hat{L}^\dagger$. 
	The case of non-Hermitian $\hat{L}$ can be easily handled along the lines of Ref.~\citenum{RiSuMoe15_034115_}.
	
	\paragraph{Non-Markovian stochastic Schr\"odinger equation and the hierarchy of pure states:}
	Within the HOPS method the reduced density operator $\rho(t)$ is obtained from
	\begin{equation}
	\label{eq:rho(t)}
	  \rho(t)=  \mathbb{E}\big\{\ket{\psi_t(Z_t^*)}\bra{\psi_t(Z_t^*)}\big\},
	\end{equation}
	where the $\ket{\psi_t(Z_t^*)}$ are vectors in the system Hilbert space that depend on  stochastic processes $Z_t$, and $\mathbb{E}[\cdots]$ denotes the average over trajectories.
	The $Z_t$ are complex valued and fulfill $\mathbb{E}[Z_t]=0$ and \footnote{With our choice of the correlation functions of the stochastic processes we follow the one of the original NMQSD derivation \cite{strunz1997pla,strunz1998pra}.	There exist other choices fore the noise-correlations which might give numerical advances e.g., for high temperature~\cite{zhao2016hierarchy,shi2016alternative}.
	We discuss this scheme in section SV of the \SI~\cite{SI}.}: $\mathbb{E}[Z_tZ_s]=0$ and $\mathbb{E}[Z_tZ_s^*]=\alpha(t-s)$.
	To obtain the HOPS, the bath-correlation function (\ref{eq:BCF}) is approximated by a sum of exponentials (which we denote as modes),
	\begin{equation}
		\label{eq:BCF_exp}
		\alpha(t)\approx\sum_{k=1}^{K}d_ke^{-\nu_k t}~~(t\ge 0),
	\end{equation}
	with complex numbers $\nu_k$. 
	Then the following hierarchy of first order differential equations can be derived \cite{eisfeld2014hierarchy},
	\begin{equation}
		\label{eq:hops_linear}
		\begin{aligned}
			\partial_t\psi_t^\Bn=&\big[-i\hat{H}_\text{S}+ \hat{L} Z_t^{*}-\sum_{k=1}^{K}n_k\nu_{k}\big]\psi_t^\Bn\\
			&+\hat{L} \sum_{k=1}^{K}\frac{d_k}{\sqrt{|d_k|}} \sqrt{n_k}\,\psi_t^{\Bn-\Be_k}\\
			&-\hat{L}^\dagger \sum_{k=1}^{K}\sqrt{|d_k|} \sqrt{n_k+1}\,\psi_t^{\Bn+\Be_k}.
		\end{aligned}
	\end{equation}
	The superscript $\Bn=\{n_{1},\cdots,n_{k},\cdots,n_{K}\}$ consists of a set of non-negative integer indices, and $\Be_{k}=\{0,\cdots,1_{k},\cdots,0\}$. 
	The initial conditions are $\psi^{\bf 0}_{t=0}=\psi_{\mathrm{ini}}$ and $\psi^{\Bn}_{t=0}=0$ for $\Bn\ne \mathbf{0}$.
	The trajectories entering Eq.~(\ref{eq:rho(t)}) are $\psi_t(Z^*_t)=\psi_t^\mathbf{0}(Z^*_t)$.

Note that compared to the original derivation of HOPS~\cite{eisfeld2014hierarchy} we have rescaled the auxiliary vectors according to 
$
		{\psi}^\Bn_t \rightarrow \big(\prod_{k=1}^{K}n_k!\abs{d_k}^{n_k}\big)^{-\frac{1}{2}}\psi^\Bn_t.
$	

For the general case of several environments and several system-bath operators as given in Eq.~(\ref{eq:H_SB}), for each $\hat{L}_j$ one obtains the terms as on the right hand side of Eq.~(\ref{eq:hops_linear}), where all $k$ dependent quantities get an additional index $j$. One now has $J$ independent processes $Z^*_{t,j}$. 
The hierarchy is now labeled by $\Bn=\{n_{11},\cdots,n_{kj},\cdots,n_{KJ}\}$.
Details are presented in section SIII of the \SI~\cite{SI}.

In practice one has to truncate the hierarchy, which can be achieved by a suitable approximation of the terms appearing in the last line of Eq.~(\ref{eq:hops_linear}). 
Possible choices are for example the `terminator' suggested in Ref.~\cite{eisfeld2014hierarchy}, or simply setting these terms to  to zero, as we do here.
To keep the number of coupled equations reasonably small proper truncation is an important issue \cite{eisfeld2018flexible}. 
For example for the common 'triangular' truncation scheme $\sum_{j=1}^J\sum_{k=1}^K n_{jk}\le \mathcal{N}_\mathrm{max}$, where $\mathcal{N}_\mathrm{max}$  determines the 'depth' of the hierarchy.
For given $J$, $K$ and $\mathcal{N}_\mathrm{max}$ the number of equations is then approximately given by $\frac{1+\mathcal{N}_\mathrm{max}}{JK}  \binom{JK+\mathcal{N}_\mathrm{max}}{1+\mathcal{N}_\mathrm{max}}$. 
This shows that even for $\mathcal{N}_\mathrm{max}=2$ the size of the hierarchy is massive, if the total number of modes ($JK$) is large.
To make things worse, for many relevant parameter regimes a large $\mathcal{N}_\mathrm{max}$
 is required.
Our MPS/MPO formulation will resolve this fundamental problem. 

%
	
	\paragraph{Effective Hamiltonian for HOPS:}

	To obtain at a convenient form to construct MPS and MPO, we formally define states
 $\{\ket{\Bn}\}$ with $\ket{\Bn}=\ket{n_{1},\cdots,n_{K}}$ and introduce
	\begin{equation}
		\label{eq:mps}
		\begin{aligned}
			\ket{\Psi(Z^*)}_t
			=&\sum_{\Bn}\psi_t^{{\Bn}}(Z^*)\ket{\Bn}
		\end{aligned}
	\end{equation}
    with the auxiliary vectors $\psi_t^{\Bn}$ of HOPS as expansion coefficients. 
   Defining the following orthonormal relation,	$\braket{\Bn|\Bn'}=\delta_{\Bn\Bn'}$, these coefficient can be obtained from $\psi_t^{{\Bn}}=\braket{\Bn|\Psi}_t$. 
The HOPS system of equations (\ref{eq:hops_linear}) is then expressed as
	\begin{equation}
		\label{eq:EOM}
		\begin{aligned}
			\partial_t\ket{\Psi(Z^*)}_t=&-i\hat{H}_{\text{eff}}(Z^*)\ket{\Psi(Z^*)}_t,\\
		\end{aligned}
	\end{equation}
	with the effective stochastic Hamiltonian
	\begin{equation}
		\label{eq:hops_h_eff_linear_rescale}
		\begin{aligned}
			\hat{H}_{\text{eff}}=&\hat{H}_\text{S}+i\hat{L} Z_t^*-i\sum_{k=1}^{K}\nu_{k}\,\hat{b}_{k}^\dagger \hat{b}_{k}\\
			&-i \hat{L}^\dagger\sum_{k=1}^{K} \sqrt{\abs{d_{k}}} \,\hat{b}_{k} +i \hat{L} \sum_{k=1}^{K}\frac{d_k}{\sqrt{\abs{d_k}}}\,\hat{b}_{k}^\dagger,
		\end{aligned}
	\end{equation}
	where  creation $(\hat{b}^\dagger_{k})$ and annihilation $(\hat{b}_{k})$ have been defined by
	\begin{equation}
		\begin{aligned}
			\hat{b}^\dagger_{k}\ket{\Bn}=&\sqrt{n_{k}+1}\ket{\Bn+\Be_{k}}\\
			\hat{b}_{k}\ket{\Bn}=&\sqrt{n_{k}}\ket{\Bn-\Be_{k}}.
		\end{aligned}
	\end{equation}
	Now the labels $\{n_k\}$ of the hierarchy play the role of \textit{occupation numbers}. 
	Thus we refer to the states $\ket{\Bn}$ as \textit{pseudo-Fock} states.
	It is worth mentioning that the hierarchy labels $\{n_k\}$ do not appear anymore in $H_{\text{eff}}$ and that the third term in the right hand side looks like a collection of harmonic oscillators, however with complex frequencies $\{\nu_k\}$.

	Before we introduce the MPS representation, let us mention that for numerical calculations, in particular for strong system bath coupling, a non-linear, normalizable version of Eqs.~(\ref{eq:hops_linear}) and~(\ref{eq:hops_h_eff_linear_rescale}) is required to achieve convergence with respect to the number of trajectories \cite{eisfeld2014hierarchy}.
	This non-linear equation is obtained through the following replacements: $\hat{L}^{\dagger}\rightarrow \hat{L}^{\dagger}-\langle\hat{L}^{\dagger} \rangle_t$  and $Z_{t}^{*}\rightarrow Z_{t}^{*}+\int_0^t\mathrm{d}s\,\alpha^{*}(t-s)\langle{\hat{L}}^{\dagger}\rangle_s$. 
Expectation values  $\langle\cdot\rangle_t$ are calculated using the normalized state.

	\paragraph{HOPS in MPS/MPO representation (HOMPS):}
	The sum-of-products form of Eq.~(\ref{eq:hops_h_eff_linear_rescale}) is convenient for an implementation in terms of MPSs and MPOs.
	We represent the wavefunction in Eq.~\eqref{eq:mps} as an MPS by expanding $\ket{\Psi}_t$ on a product of system  states $(\ket{\ell})$ and \textit{pseudo-Fock} states, that is $\ket{\ell}\otimes\ket{\Bn}$,
	\begin{equation}
		\label{eq:mps_sb}
		\begin{aligned}
			\ket{\Psi}_t &=\sum_{\ell,\Bn}{\psi}_t^{{\ell,\Bn}}\ket{\ell,n_{1}, \cdots, n_{K}} \\
			& = \sum_{\ell,\Bn,\mathbf{a}} A^\ell_{1, a_0} A^{n_1}_{a_0 a_1} \cdots A^{n_K}_{a_{K-1,1}} \ket{\ell,n_{1}, \cdots, n_{K}}.
		\end{aligned}
	\end{equation}
	Each $A^{n_i}_{a_{i-1} a_i}$ is a rank-3 tensor with 'physical index' $n_i$ and 'virtual indices' $a_{i-1}$ and $a_i$.  
	The ranges of the virtual indices are denoted as bond dimensions $M_i$.
	Increasing the bond dimensions can  systematically improve the accuracy of an MPS.
	For fixed bond dimensions, the computational cost to evolve Eq.~\eqref{eq:EOM} is polynomial rather than exponential with the number of effective modes $K$.
   The bond dimensions can also be optimized adaptively in each time-step during propagation.
   We present calculations using both methods in this Letter and the \SI~\cite{SI}.
   Note, that the physical dimensions $n_k$ stem from the original indexing of the HOPS.
  As long as the bond-dimensions are not too large, one can go to large $\mathcal{N}_\mathrm{max}$ without a drastic increase of the MPS representation. One can even set the same maximal value $n_\mathrm{max}$ for all modes. 
  For the case of Eq.~(\ref{eq:H_SB}) this would correspond to $n_\mathrm{max}^{JK}$ coupled equations in HOPS. 
	In conjunction with the MPS,
    Eq.~\eqref{eq:hops_h_eff_linear_rescale} is written as a MPO, 
    \begin{equation}
	\label{eq:mpo}
	\begin{aligned}
	    \hat{H}_\mathrm{eff}  =\sum_{\ell, \ell', \Bn, \Bn',\mathbf{w}} & W^{\ell,\ell'}_{1,w_0} W^{n_1,n_1'}_{w_0 w_1} \cdots W^{n_K,n'_K}_{w_{K-1,1}}  \\
	    & \ket{\ell,n_{1}, \cdots, n_{K}} \bra{n'_{K}, \cdots, n'_{1},\ell'},
	\end{aligned}
	\end{equation}
	in which $W^{n_i,n'_i}_{w_{i-1}w_i}$ is a matrix of local operators acting only on the $i$th effective mode. 
	This factorization is not unique. 
	We adopt the bipartite graph based algorithm~\cite{ren2020mpo} to construct the most compact MPO with the smallest size of virtual indices $w_i$, in order to reduce the computational cost of tensor contractions. 
    Introducting MPOs allows one to calculate $\hat{H}_\mathrm{eff} \ket{\Psi}_t$ using contractions of local matrices, which is then of polynomial complexity.
    We stress that all tensors $A$ and $W$ depend on the stochastic processes.

	The generalization to more than one environment as given in Eq.~(\ref{eq:H_SB}) is straight forward (see the section SIII of the \SI~\cite{SI}).
	We would like to emphasize that ordering (and to some extend also the number) of the tensors in MPS can be chosen according to the specific form of the system Hamiltonian $H_\mathrm{S}$ and the coupling operators $H_{\mathrm{SB},j}$, as we will examplify below.

	\begin{figure}
		\includegraphics[width=8.6 cm]{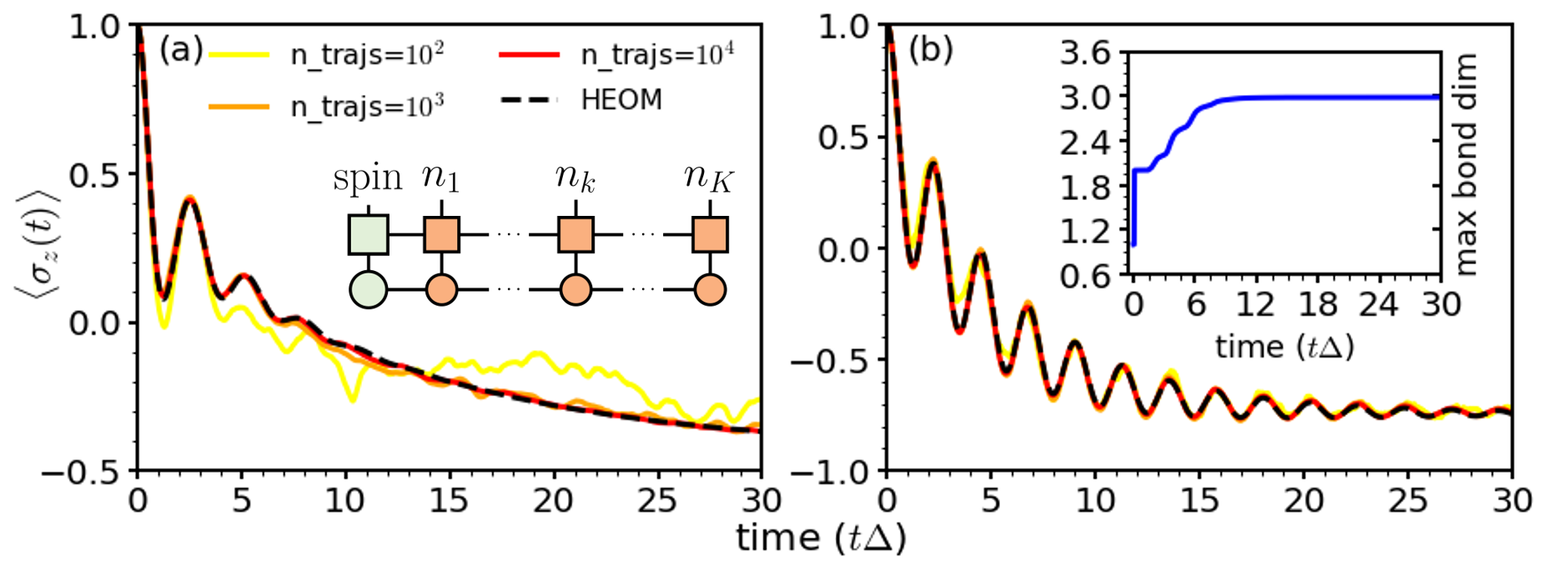}
		\caption{\label{fig2}  Population dynamics of the spin-boson model with $\epsilon=1.0,~\Delta=1.0,~\eta=0.5$  by averaging over $10^2$ (yellow), $10^3$ (orange) and $10^4$ (red) trajectories.   
		(a) High temperature $\beta=0.5$ and small $\gamma=0.25$.
		HOMPS results are obtained using $K=1$ and  $n_\mathrm{max}=39$.
		As inset the MPS/MPO arrangement is shown, which is used for all SBM calculations.
		(b) Low temperature, large damping case, with $\gamma=5.0, \text{and}~\beta=50.0$. HOMPS results with $K=13$, and $n_\mathrm{max}=9$ for each mode. \textit{Inset}: Evolution of maximum bond dimension averaged over $10^3$ trajectories.
		The HEOM  results (black, dashed) are taken from Ref.~\cite{shi2016alternative}. 
		}
	\end{figure}

	\paragraph{Numerical example 1: the Spin-Boson model (SBM).}
		The SBM is often used to test the applicability of a new method. Here $\hat{H}_\mathrm{S}=\epsilon\sigma_z+\Delta\sigma_x$ and  $\hat{L}=\sigma_z$, where  $\sigma_x=\ket{1}\bra{2}+\ket{2}\bra{1}$ and $\sigma_z=\ket{1}\bra{1}-\ket{2}\bra{2}$. 
	We consider a Debye spectral density 
	$
		S(\omega)=\eta\frac{\omega\gamma}{\omega^2+\gamma^2}.
	$
	In Ref.~\cite{shi2016alternative} calculations using the density matrix based HEOM method have been presented for 
	$\epsilon=1.0$, $\Delta=1.0$ and $\eta=0.5$ and (a) a `high temperature low damping' case with $\gamma=0.25$ and $\beta=0.5$; (b) a `low temperature large damping' case with $\beta=50$ and $\gamma=5.0$. 
	In Fig.~\ref{fig2} we show that HOMPS quickly converges to these reference calculations. 
	For the high temperature case, panel (a), 1000 trajectories give very good agreement, for the low temperature case, panel (b), where fluctuations of the noise are smaller, only 100 trajectories are needed.
	This demonstrates the validity of our procedure.
	Let us now consider in more detail the complexity of the equations to solve.
	In each case we have chosen the number of modes $K$ large enough to guarantee convergence of the bath-correlation function.
	For the high temperature case only one mode is necessary ($K=1$), while for the low temperature case we used $K=13$.
For simplicity,  we use for each mode $k$ the same truncation condition  $n_{k} \le n_\mathrm{max}$. 
Although for the high temperature case we need $n_\mathrm {max}\approx 40$ (see section SVI of the \SI~\cite{SI}), the problem is still small, because of $K=1$.
The low temperature case with $K=13$ is more challenging.
While for HOPS this would results in $9^{13}$ equations and one would have to use adequate  truncation procedures~\cite{eisfeld2018flexible}, for HOMPS it only means a small increase computational effort.
Relevant for HOMPS is the size of the tensors in Eq.~(\ref{eq:mps_sb}), which is given by the product of $n_\mathrm{max}$ and the two bond dimensions.
Remarkably, as shown in the inset of Fig~\ref{fig2}(b), the actual maximum bond dimension $M_\textrm{max}$  is almost always smaller than 3.
This means that the largest tensor has a dimension around $3\cdot 3\cdot n_\mathrm{max}=81$, for the used $n_\mathrm{max}=9$.

\paragraph{Numerical example 2: Exciton transport in a linear chain.}

\begin{figure}
    \centering
    \includegraphics[width=8.6 cm]{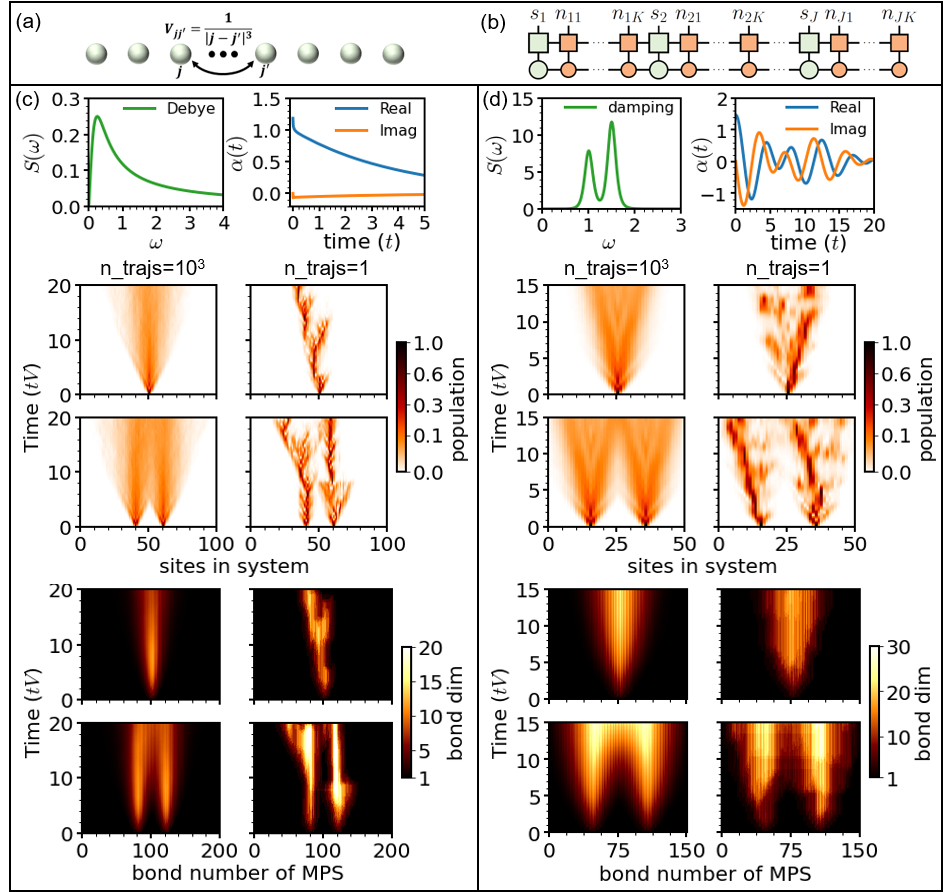}
		\caption{\label{fig:Holstein} Application of HOMPS to a linear chain. 
		(a) one dimensional chain with intermolecular coupling strength $V_{jj'}=\frac{1}{|j-j'|^3}$. 
		(b) The used MPS/MPO structure. 
		(c), (d):
		Evolution of population and bond dimension for both average and single  trajectories.
		The corresponding spectral densities and bath-correlation functions are shown in the upper row of each panel.
		Parameters of HOMPS are (c) $K=1$ and $n_\mathrm{max}=40$. 
		(d)  $K=2$, $n_\mathrm{max}=20$.
		} 	
    \end{figure}

As a second example we consider the motion of (electronic) excitations under the influence of damped vibrational modes. 
Such a model describes e.g.\ molecular aggregates  or biological light harvesting systems with coupling to  vibrations of the molecules \cite{VaEiAs12_224103_,jang2018delocalized}.
Treating each molecule as an electronic two-level system, the total Hamiltonian can be written as $\hat{H}=\sum_{j=1}\hat{H}_j+ \sum_{jj'}\hat{V}_{jj'}$, where the Hamiltonian $\hat{H}_j$ of the $j$th site is characterized by a system part $\hat{H}_{\mathrm{S},j}=\epsilon_j a^{\dagger}_ja_j$, system-bath coupling operators $\hat{L}_j=a^{\dagger}_ja_j$ and a corresponding spectral density $S_j(\omega)$, which contains molecular vibrations and the coupling to the local surroundings. 
	The coupling between sites is typically the long-range dipole-dipole interaction and assumed not to couple directly to the environment.
	Further details are given in the section SIV of the \SI~\cite{SI}.
	For this problem we use a MPO/MPS as shown in Fig.~\ref{fig:Holstein}b, where each local system Hamiltonian $H_{\mathrm{S},j}$ is followed by its modes from the decomposition of the respective bath-correlation function.
This allows us to also conveniently treat the case of several electronic excitations, needed for example to describe exciton-exciton annihilation experiments \cite{RyKoMa01_5322_}. 
In Fig.~\ref{fig:Holstein} we show electronic excitation transport along an one-dimensional chain with nearest neighbor interaction $V_{jj'}=V\, \delta_{jj'}$.
A case of long-range dipole-dipole interaction with $V_{jj'}=1/|j-j'|^3$ is presented in the \SI~\cite{SI}.  
In Fig.~\ref{fig:Holstein} we present results for two different spectral densities and temperature regimes. 
In panel (c) we use a Debye spectral density with the same parameters as for the high temperature case of the SBM (cf.~Fig.~\ref{fig2}a). 
In panel (d) the spectral density consists of two broadened peaks; a spectral density typical for weakly damped vibrational modes of polyatomic molecules.
Here we consider zero temperature.
These spectral densities and the corresponding bath-correlation functions are shown in the upper row for both cases.
Below, we show the time dependent populations, for the converged results and for single trajectories and on the bottom the time dependence of the bond dimensions.
Additional examples of single trajectories can be found in the \SI~\cite{SI}.
We see that in both cases the bond dimensions remain small and relatively well localized.
This is an additional benefit for handling such large systems.

	\paragraph{Conclusions:}
	\label{sec:summary}
	
	The numerical results demonstrate that HOMPS works well in simulating quantum dissipative dynamics for large systems in highly non-Markovian regimes.  
	This is achieved by a MPS/MPO representation of HOPS.
	Compared to Ref.~\cite{flannigan2021many}, which parallels our work, our focus has been the treatment of several modes per site, which is important, e.g., when treating temperature or vibrational modes of molecules.

	We have used here a representation of HOPS where the hierarchy is constructed from an decomposition of a bath-correlation function that contains the temperature.
	Recently Hartmann and Strunz have derived a version of HOPS where the temperature enters simply as a {\it classical} stochastic process, and the hierarchy is constructed from the zero-temperature bath-correlation function ~\cite{strunz2017exact}.
	This approach can also be readily used within the MPS/MPO of the present work.

   An appealing feature of the present HOMPS is that the reduction in size can be done by automatically adapting the bond-dimensions in each time step.
In that sense HOMPS shares similarities to other adaptive schemes for Markovian and non-Markovian quantum state diffusion~\cite{gao2019charge,doran2021adaptive}.
A promising future direction is to meld HOMPS with such schemes. 
	We believe that HOMPS is a fruitful approach  to explore the dissipative dynamics in open quantum systems.

	\vspace{0.9cm}

	\begin{acknowledgements}
	    We thank Qiang Shi for help on the high temperature approach to HOPS. We thank C.~W\"achtler and Jiushu Shao for reading the manuscript. 
		X.G.~acknowledge support from Sun Yat-sen University "100 Top Talents Program", Startup Grant and computational resources and services provided by national supercomputer center in Guangzhou. 
		Z.S. and J.R. acknowledge support from the National Natural Science Foundation of China
        (NSFC), Grant Number 21788102 and 22003029, as well as from the Ministry of Science and Technology of China through the National Key R\&D Plan, Grant
        Number 2017YFA0204501.
        A.E. acknowledges support from the DFG via a Heisenberg fellowship (Grant No EI 872/5-1).
	\end{acknowledgements}

	\bibliographystyle{apsrev4-2}

\end{document}